\documentclass{IEEEtran}
\IEEEoverridecommandlockouts
\usepackage{cite}
\usepackage{tabularx}
\usepackage{amsmath,amssymb,amsfonts}
\usepackage{algorithmic}
\usepackage{multirow}
\usepackage{ragged2e,array,booktabs}
\usepackage{graphicx}
\usepackage{textcomp}
\usepackage{xcolor}
\usepackage{hyperref}
\usepackage{bookmark}
\usepackage{subfigure}
\usepackage{float} 
\def\BibTeX{{\rm B\kern-.05em{\sc i\kern-.025em b}\kern-.08em
		T\kern-.1667em\lower.7ex\hbox{E}\kern-.125emX}}

\makeatletter
\def\endthebibliography{%
	\def\@noitemerr{\@latex@warning{Empty `thebibliography' environment}}%
	\endlist
}
\makeatother
\begin{document}	
\title{Multiple-type Transmission Multiple-type Reception Framework on Molecular Communication}
\author{Hamid Khoshfekr Rudsari, Mohammad Reza Javan,~\IEEEmembership{Member,~IEEE}, Mahdi Orooji,~\IEEEmembership{Member,~IEEE},
	Nader Mokari,~\IEEEmembership{Senior Member,~IEEE}, and
	Eduard A. Jorswieck,~\IEEEmembership{Senior Member,~IEEE}
	\thanks{H. K. Rudsari, M. Orooji, and N. Mokari are with the Department of Electrical and Computer Engineering, Tarbiat Modares University, Tehran, Iran. email: \{hamid\_khoshfekr, morooji, nader.mokari\}@modares.ac.ir}	
	\thanks{M. R. Javan is with the Department of Electrical and Robotic Engineering, Shahrood University of Technology, Shahrood, Iran. e-mail: javan1378@yahoo.com}
	\thanks{E. A. Jorswieck is with the Department of Information Theory and Communication Systems, TU Braunschweig, Germany. e-mail: jorswieck@ifn.ing.tu-bs.de}}
\markboth{IEEE Communications Letters}%
{}
\maketitle
\begin{abstract}
In this paper, we propose a new Multiple-Input Multiple-Output (MIMO) Molecular Communication (MC) system where multiple types of molecules are utilized for transmission and reception of information. 
We call the proposed framework as Multiple-type Transmission and Multiple-type Reception (MTMR). We also obtain the bit error rate (BER) of the system and an optimization problem is formulated to minimize BER by optimizing the drug dosage for designing drug release mechanism. As numerical analysis shows, the BER of MIMO-MTMR in MC is minimized to $\text{3.7}\times\text{10}^{\text{-3}}$ by considering the budget of molecules as 10000. Furthermore, MIMO-MTMR outperforms Single-type Transmission Single-type Reception MIMO from the BER performance point of view approximately 54\% for time slot 10s.
\end{abstract}
\section{Introduction} \label{sec:introduction}
\IEEEPARstart{M}{olecular} Communication~(MC) is a promising paradigm which can hold the operation of communication between nanomachines where they send molecules to convey information~\cite{nakano2013molecular,akyildiz2008nanonetworks}.

The transmitters in prior works on Multiple-Input Multiple-Output~(MIMO)-based MC release one type of molecules in order to send information~\cite{koo2015detection,koo2016molecular,damrath2017spatial,zabini2018molecular}. We call this type of MIMO as Single-type Transmission Single-type Reception~(STSR). In~\cite{koo2015detection}, the authors propose three algorithms for detection of the output bits which are based on the estimated channel response. 
In~\cite{koo2016molecular}, the authors provide five detection algorithms which require some information about the channel and the topology of the system. In spite of that, MC systems can not bear this complexity~\cite{noel2015joint}. In this regard, designing a new framework for MIMO-based MC can be beneficial in terms of the complexity reduction of the system.

In this paper, we propose a molecular communication via diffusion~(MCvD) system based on MIMO in which multiple types of molecules are utilized for transmission information. Utilizing MIMO in MC can significantly increase the data rate, due to the fact that more number of bits of information are sent toward the receiver in a time interval. On this subject, we propose a new framework on MCvD where the transmitter nanomachines release different types of molecules to convey information. We name the proposed MIMO-based MCvD as Multiple-type Transmission Multiple-type Reception~(MTMR). The proposed framework is beneficial in such drug delivery systems~(DDSs) that aim at releasing different types of drugs toward the target cells. We assume that each target cell is sensitive to one specific drug. To understand the target cells reaction by absorbing the drugs, we model them as spherical absorbing receivers \cite{cao2019diffusive}. The transmitter nanomachines are assumed as point sources~\cite{guo2016molecular}. The considered channel model is an unbounded 3-dimensional~(3D) diffusive environment. In the regard of evaluating how the communication process is continuing without error, we consider Bit Error Rate (BER) metric. It could be implemented in networks of nanomachines that diagnosing and treatmenting the specified objects such as cancer cells are the main purpose of DDSs \cite{akan2016fundamentals}. We also formulate an optimization problem to find the optimized number of molecules allocated to each transmitter nanomachine regarding the minimization of BER in case of utilizing MIMO-based MCvD. From the DDS point of view, we optimize the drugs' dosages to minimize the error probability of delivering the drugs toward the target cells.

The remainder of the paper is organized as follows. In Section~\ref{sec:system_model}, we investigate the system model in details. The different structures of communication as single-input single-output~(SISO), single-input multiple-output~(SIMO), multiple-input single-output~(MISO), and MIMO are assessed in Section~\ref{sec:structures}. The numerical analysis is given in Section~\ref{sec:numerical_result} and the paper is concluded in Section~\ref{sec:conclusion}. 
\section{System Model} \label{sec:system_model}
\begin{figure}[]
	\centering
	\scalebox{.1}{}
	\includegraphics[width=230pt]{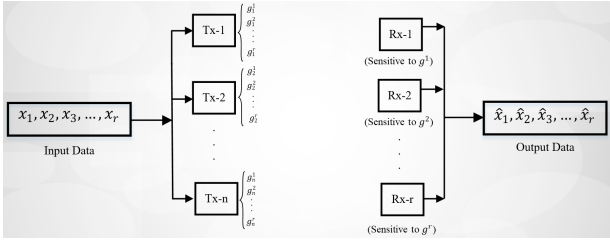}
	\caption{The illustration of MIMO-MTMR MCvD to design the drug releasing mechanism.}
	\label{Fig:Block}
\end{figure}

We consider a MC framework handling the behavior of a diffusive environment. We employ $n$ nanomachines where they cooperate with each other to release drugs into the desired locations. The cooperation process is based on two parameters: I)~time of releasing the molecules, i.e., the time slot duration, and II)~the drug dosage, i.e., the number of molecules. In this context, we illustrate the desired systematic scheme of a MIMO-based MCvD system with cooperative nanomachines in Fig.~\ref{Fig:Block}. We consider cooperating nanomachines in which they have multi-layer liposome and can release the molecules of different types~\cite{chude2019nanosystems}. Then, the input data (see Fig.~\ref{Fig:Block}) has the length of $r$, i.e., we aim to transmit $r$ bits. Each bit is conveyed with one type of molecules, therefore, each transmitter emits $r$ types of molecules into the environment. The molecules are propagated into the medium until they arrive at the receivers and absorbed by them. The receivers are nanomachines each of which is sensitive to a specific type of molecules~\cite{moy1994intermolecular}. As an instance, the transmitters release $\{1, 2, ..., r\}$ types of molecules to send information bits as $\{x_1, x_2, ..., x_r\}$ where $x_i = \{0, 1\}$ for $i = 1, 2, ..., r$.  We also consider a 3D unbounded diffusive environment. The diffusion coefficient of the molecule of type~$\theta \in \{D_1, D_2, ..., D_r\}$ in the diffusive fluid environment is related to the temperature of the fluid, the dynamic viscosity of the medium, and the Stoke's radius of the molecules of type~$\theta$~\cite{tyrrell2013diffusion}.


The molecules propagate through the medium via the Brownian motion~\cite{mori1965transport}. The probability density function (PDF) of the absorption time $t$ by the spherical receiver Rx-k for the molecule of type $\theta$ which is transmitted from Tx-s is \cite{yilmaz2014three}
\begin{align}
	\gamma_{k,\theta}^s (t) = \dfrac{r_k d^s_k}{(d^s_k + r_k) \sqrt{4 D_\theta t^3}} \exp(-\dfrac{d^{s^2}_k}{4 D_\theta t}), \label{eq:pdf}
\end{align}
where $r_k$ and $d^s_k$ are the radius of the receiver Rx-k and the distance between transmitter Tx-s and the surface of the receiver Rx-k, respectively. In addition, $\exp(.)$ is the exponential function. It is worth noting that the receivers fully absorb the molecules and count them. The probability of hitting the receiver Rx-k within time $t$ by the molecule of type~$\theta$ released from transmitter Tx-s is the cumulative distribution function (CDF) of (\ref{eq:pdf}). To attain CDF, we should integrate (\ref{eq:pdf}) from $t - t_0$ (the releasing time) to $t$~\cite{papoulis2002probability}. Therefore, the probability of reception the molecule of type $\theta$ by receiver Rx-k which is released from Tx-s is given by
\begin{align}
	\Gamma_{k,\theta}^s (t) =  \int_{t - t_0}^{t} \gamma_{k,\theta}^s(u)\, du. \label{eq:CDF}
\end{align}
To simplify (\ref{eq:CDF}), first, we consider  $t_0 = 0$, next, we denote the probability reception of the molecule of type $\theta$ released from the transmitter Tx-s as $\Gamma_{k,\theta}^s (t) = P_{\text{rec}} (D_\theta, d^s_k, r_k, t)$.

\subsection{Biological Aspects} \label{subsec:Biological}
\begin{figure}[t]
	\centering
	\scalebox{.1}{}
	\includegraphics[width=200pt]{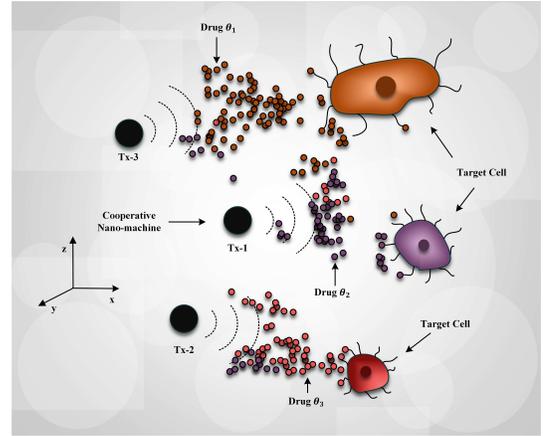}
	\caption{The Drug Release Mechanism for novel DDS to deliver the drugs from three cooperating nanomachines to three different target cells.}
	\label{Fig:Env}
\end{figure}

The basis of the proposed model is to deliver the drugs released from the transmitters, i.e., the cooperating nanomachines, to the receivers, i.e., the target cells. The proposed scheme to deliver the drugs to the target cells is provided in Fig.~\ref{Fig:Env}. Each transmitter nanomachine releases different types of molecules (drugs) into the environment where it is desired to be absorbed by the target cells. We assume the target cells as the fully absorbing spherical receivers. The receivers reception process is based on ligand receptors. The receptors in biological science are divided into two categories, namely as intra-cellular receptors and cell-surface receptors~\cite{moy1994intermolecular}. The intra-cellular receptors take place inside the cell, such as nucleus or cytoplasm and the molecules that can be absorbed by them should cross the plasma membrane to reach them. The cell-surface receptors are anchored to the membrane of the cell and the many types of molecules can be absorbed by them without crossing the plasma membrane. In this paper, we consider intra-cellular receptors as the receivers in the proposed MIMO-based MCvD system. The reason behind this is the aforementioned type of receptors can sense less types of molecules~\cite{moy1994intermolecular}. Therefore, they are suitable to consider as receivers.
 
\subsection{Molecular Communication Aspects} \label{subsec:MC} 
In this subsection, we aim to study the different structures of MCvD by considering the activation of transmitter and receiver nanomachines. We consider the number of molecules of type $\theta$ allocated to transmitter Tx-s as $g_s^\theta$ which is illustrated in Fig.~\ref{Fig:Block}. It is also assumed that the transmitters are point sources \cite{heren2015effect}, and hence, they have no volume. The input sequence is composed of $r$ bits and each bit of information leads the MCvD to utilize one type of molecules. The structure of the MCvD system can handle SISO, SIMO, MISO, and MIMO. In the following section, we determine the different structures of the MCvD system via the proposed model.

\section{Multiple Transmission in Molecular Communication}\label{sec:structures}
In this section, we introduce the procedures of SISO, SIMO, MISO, and MIMO to convey information from the nanomachine transmitters to the nanomachine receivers. Since, the BER is an important criterion to evaluate the effectiveness of the drug delivery~\cite{chahibi2017molecular}, we calculate this metric for each procedure. In the following, we assess the aforementioned procedures to attain the distribution of the number of the received molecules and BER.

\subsection{Single-input Single-output} \label{subsec:SISO}
By employing this procedure, we have one transmitter node and one receiver node. The input data has the length of $r$ and one type of molecules~(namely type $\theta$) is utilized. Each bit of the input data is transmitted in one time slot in SISO-based MCvD. The number of molecules which are transmitted in the current time slot $m$ is denoted by $N^\theta[m]$. We also utilize the on-off keying (OOK) modulation according to its efficient reception probability~\cite{garralda2011diffusion}. In this context, the transmitter releases $g^\theta$ molecules in order to send bit ``1'' and no molecules to send bit ``0''. The distribution of the number of the received molecules is binomial due to the fact that the movements of them are independent of each other~\cite{kuran2011modulation}. Therefore, the distribution of the received molecules at the current time slot $m$ is
\begin{align}
	N^{\text{SISO}}_{\text{cur}} [m] \sim \mathcal{B}\big( x[m] g^\theta, p^\theta_{s,k} \big), \label{eq:current_bino_SISO}
\end{align}
where $\mathcal{B}(.,.)$ is the binomial distribution~\cite{papoulis2002probability}, $x[m]\in\{0,1\}$ is the bit transmitted by transmitter Tx-s at time slot~$m$, and $p^\theta_{s,k} = P_{\text{rec}} (D_\theta, d^s_k, r_k, t)$. Furthermore, there are some molecules released from previous time slots, but, received in the current time slot $m$. They are considered as Inter-symbol Interference (ISI). We assume the number of previous time slots which are involved in ISI, i.e., the ISI length, are limited~\cite{kilinc2013receiver}. Therefore, the distribution of the molecules of type $\theta$ named as ISI is given by
\begin{align}
	N^{\text{SISO}}_{\text{ISI}}[m] \sim \sum_{j=1}^{J} \mathcal{B} \big( x[m-j] g^\theta, p^{\theta,j+1}_{s,k} - p^{\theta,j}_{s,k}\big), \label{eq:ISI_bino_SISO}
\end{align}
where $x[m-j]$ is the bit transmitted from the $j^{\text{th}}$ previous time slot and $p^{\theta,j}_{s,k} = P_{\text{rec}} (D_\theta, d^s_k, r_k, jt)$. After some mathematical manipulations and by approximating the binomial distribution to Normal distribution, the number of molecules of type $\theta$ which are transmitted by Tx-s at the beginning of time slot $m$ and received by Rx-k at the end of time slot $m$ are given by~\cite{tavakkoli2017performance}
\begin{align}
\begin{split}
	N^\theta_{s,k}&[m]  \\  \sim  & \ \mathcal{N} \bigg( x[m] g^\theta p^\theta_{s,k} \ , \ x[m] g^\theta p^\theta_{s,k}(1-p^\theta_{s,k})  \bigg) \\&+ \sum_{j=1}^{J} \mathcal{N}  \bigg(  x[m-j] g^\theta q^{\theta,j}_{s,k} \ , \ x[m-j] g^\theta q^{\theta,j}_{s,k} (1-q^{\theta,j}_{s,k})  \bigg), \label{eq:received_SISO}
	\end{split}
\end{align}
where $\mathcal{N}\big(.,.\big)$ is the normal distribution and $ q^{\theta,j}_{s,k} = p^{\theta,j+1}_{s,k} - p^{\theta,j}_{s,k}$. Therefore, the number of the received molecules follows the normal distribution as
\begin{subequations}
\begin{align}
\begin{split}
\text{Pr}(N^\theta_{s,k}[m] \mid x[m] = 0) \sim \mathcal{N}(a^{0,\theta}_{s,k},b^{0,\theta}_{s,k}), \label{eq:x_0_SISO}
\end{split}
\\
\begin{split}
\text{Pr}(N^\theta_{s,k}[m] \mid x[m] = 1) \sim \mathcal{N}(a^{1,\theta}_{s,k},b^{1,\theta}_{s,k}), \label{eq:x_1_SISO}
\end{split}
\end{align}
\end{subequations}
where $a^{0,\theta}_{s,k}$ and $a^{1,\theta}_{s,k}$ are the means of the number of molecules of type $\theta$ released from Tx-s and received by Rx-k when bit ``0'' and ``1'' are transmitted, respectively. In addition, $b^{0,\theta}_{s,k}$, and $b^{1,\theta}_{s,k}$ are the variances of the number of molecules of type $\theta$ released from Tx-s and received by Rx-k when bit ``0'' and ``1'' are transmitted, respectively.
The means and variances are calculated from (\ref{eq:received_SISO})~\cite{tavakkoli2017performance}, and is different for each type of molecules. The detection process at Rx-k is handled by maximum-a-posterior (MAP). In this context, Rx-k detects the bit as ``1'' if the number of received molecules are more than the calculated threshold and otherwise as ``0''. Hence, the detected bit $\hat{x}[m]$ at the receiver Rx-k is
\begin{align}
	\hat{x}[m] = \begin{cases} \label{eq:map_SISO}
	1 & \mbox{if } N^\theta_{s,k}[m] \mbox{  $ \ge \tau^\theta, $} \  \\0 & \mbox{if } N^\theta_{s,k}[m]  \mbox{  $< \tau^\theta $} , 
	\end{cases}
\end{align}
where $\tau^\theta$ is the threshold at Rx-k. Finally, by considering that the probability of sending bit ``0'' and ``1'' are identical as $\dfrac{1}{2}$, the BER of the considered MCvD system is given by~\cite{rudsari2019non}
\begin{align}
  P^{\theta}_{s,k}[m]=\frac{1}{2}+\frac{1}{4}\bigg[\text{erf}\bigg(\frac{\tau^\theta-a^{1,\theta}_{s,k}}{\sqrt{2b^{1,\theta}_{s,k}}}\bigg)-\text{erf}\bigg(\frac{\tau^\theta-a^{0,\theta}_{s,k}}{\sqrt{2b^{0,\theta}_{s,k}}}\bigg)\bigg]. \label{eq:ber_SISO}
\end{align}

	\begin{figure*}
	\centering
	\subfigure[]{%
		\label{Fig:BER}%
		\includegraphics[height=170pt]{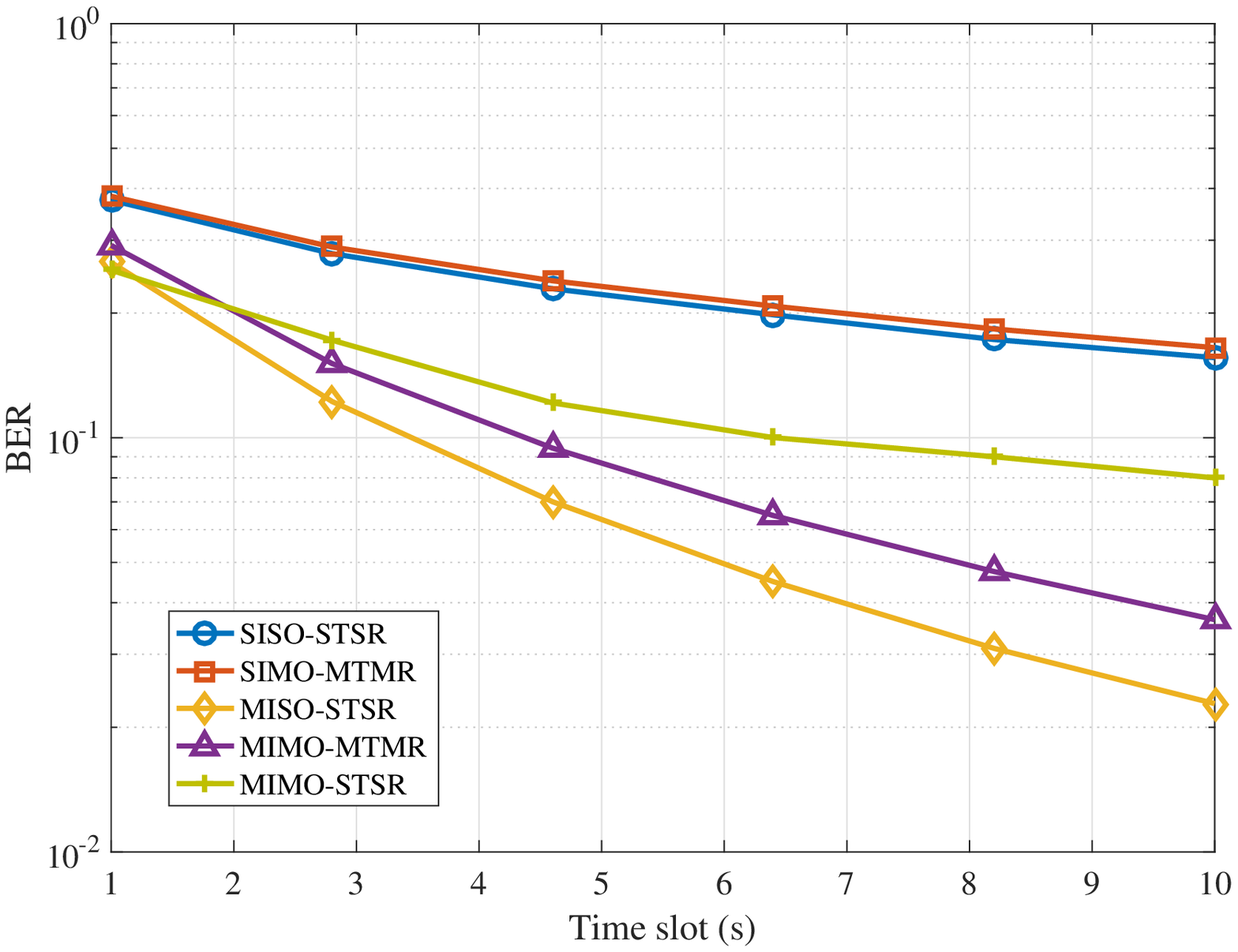}}%
	\subfigure[]{%
		\label{Fig:Opt_BER}%
		\includegraphics[height=170pt]{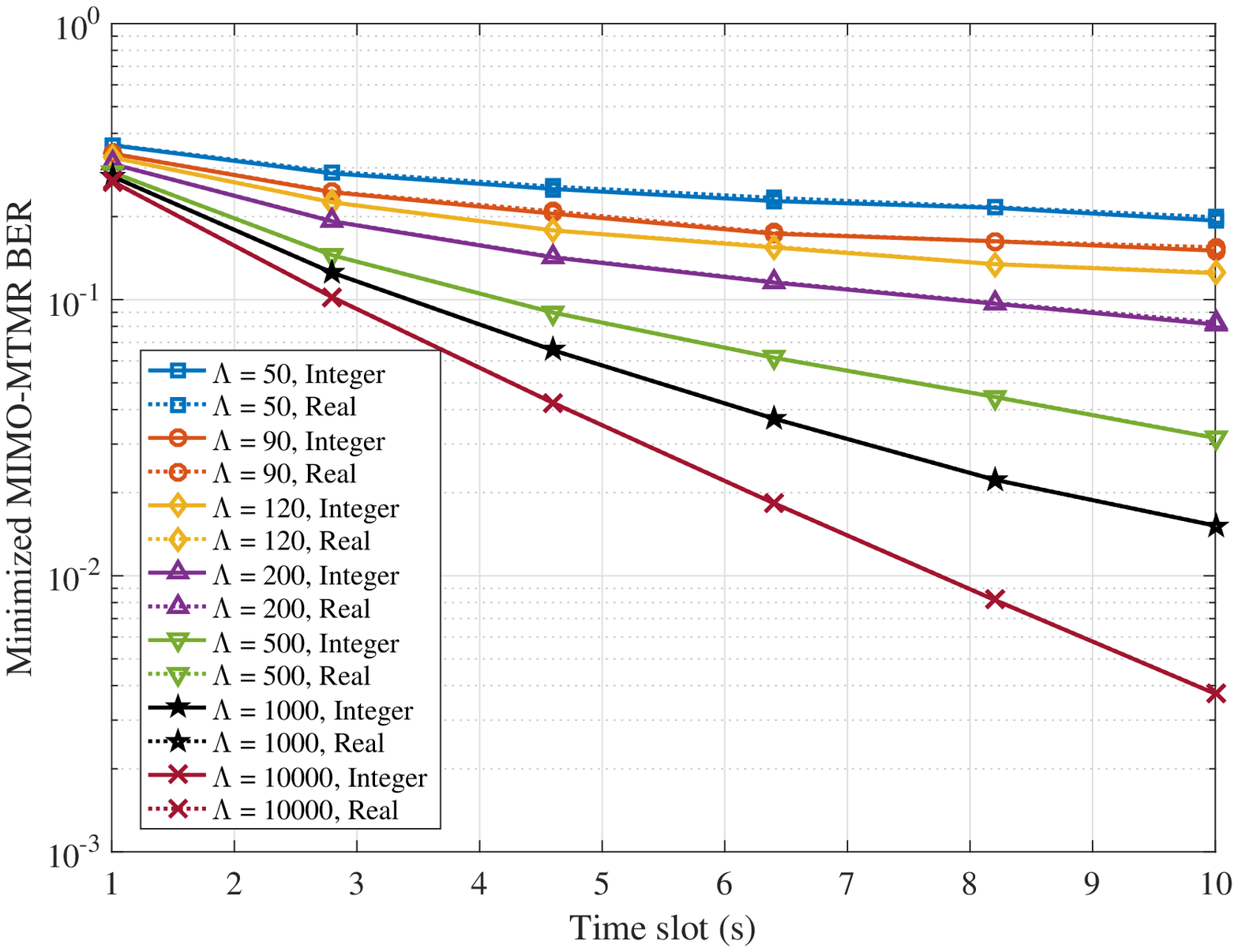}}%
	\caption{(a): The BER of the SISO, SIMO, MISO, MIMO-MTMR, and MIMO-STSR for MCvD system as a function of time slot (The number of allocated molecules to each transmitter is 1000). (b): The BER performance of the optimization problem on MIMO-MTMR as a function of time slot for different budget of molecules and considering number of molecules as real and integer variables.  }
\end{figure*}

\subsection{Single-input Multiple-output} \label{subsec:SIMO}
SIMO is the case that one nanomachine transmitter releases different drugs into the environment to be sensed by the target cells. From the MC aspect, we consider the input information as a sequence with the length of $r$, which is the same as the number of receivers. In this regard, Tx-s releases different molecules concurrently each type of them conveying one bit of information, i.e., $r$ bits are transmitted in one time slot. At the receiver side, we have $r$ receiver nanomachines each of which sensitive to the particular type of molecules.

The number of molecules released from TX-s at the beginning of the $m^{\text{th}}$ time slot, and received by Rx-k~(sensitive to the molecules of type $\theta$) at the end of the $m^{\text{th}}$ time slot is similar to (\ref{eq:received_SISO}). However, the BER calculated for each of the received bits is different because each receiver is sensitive to specific type of molecules. It is worth noting that in the proposed MCvD system, the Inter-link Interference~(ILI) does not exist, due to the fact that each receiver is sensitive to just one type of molecules~\cite{jiang2015nanoscale}. In spite of that, the STSR-based MCvD system impressed by ILI. The BER of the proposed SIMO-based MTMR MCvD system is derived as
\begin{align}
	P_{\text{SIMO}}[m] = \sum_{k = 1}^{r} P^{\theta}_{s,k}[m] \pi_k, \label{eq:ber_SIMO}
\end{align}
 where $\pi_k$ is the prior probability that bit ``1'' or ``0'' is transmitted. Note that $\pi_k$ is identical for bit ``1'' and ``0''.
\subsection{Multiple-input Single-output}\label{subsec:MISO}
From the MISO point of view, we have $n$ cooperating nanomachine transmitters in which they send a sequence of bits. The type of the releasing molecules are the same due to the existing one receiver, e.g., target cell, and therefore, the proposed MISO MCvD system is STSR-based. The transmitters release the molecules simultaneously at the beginning of the $m^{\text{th}}$ time slot to transmit one bit of information and the receiver decodes the transmitted message at the end of the $m^{\text{th}}$ time slot. The number of received molecules to Rx-k which are transmitted from Tx-1, Tx-2, ..., Tx-n, are given by
\begin{align}
N^\theta_{\text{MISO}}[m] \sim \sum_{s=1}^{n} 	N^\theta_{s,k}[m]. \label{eq:number_MISO}
\end{align}
The BER of MISO is similar to (\ref{eq:ber_SISO}), but, in spite of that, the number of received molecules in each time slot is different compared to SISO. And we name the BER of MISO as $P^\theta_{k}$ due to the fact that the transmitter nanomachines cooperate with each other to transmit one bit of information.

\subsection{Multiple-input Multiple-output} \label{subsec:MIMO}
In this subsection, we aim to find the BER of the proposed framework on MIMO-based MCvD. The transmitters send $r$ bits at the beginning of the $m^\text{th}$ time slot and the receiver nanomachines receive them at the end of the $m^\text{th}$ time slot. Therefore, the information rate of the system is in the highest performance among SISO and MISO structures and equivalent to SIMO. In the MIMO context, the cooperating nanomachines can release different types of drugs toward the target cells which are sensitive to different drugs, and hence, the proposed MIMO-MTMR can help in such novel DDS that are developed to cure the diseases based on MC \cite{chahibi2017molecular}.

The number of received molecules of type $\theta$ at Rx-k is similar to (\ref{eq:number_MISO}). However, the BER of MIMO-MTMR is not similar to MISO. The BER of MIMO-based MCvD is
\begin{align}
P^{\text{MTMR}}_{\text{MIMO}}[m] = \sum_{k=1}^{r} P^\theta_{k} \pi_k. \label{eq:ber_MIMO}
\end{align}

To design a drug release mechanism on MIMO-MTMR based MCvD system, we optimize the number of molecules~(drug dosage) by minimizing the error probability. In this regard, the optimization problem is formulated as follows
\begin{subequations}\label{eq:STDN}
\begin{align}
\begin{split}
	&\min_{\textbf{G}} \   P^{\text{MTMR}}_{\text{MIMO}}[m]
	\end{split}
	\\
	\begin{split}\label{eq:STDN_3}
	&\text{s.t.~:} \sum_{i=1}^{r} g^{\theta_i}_s = \Lambda,  \ \  \   s=1, 2,..., n,
	\end{split}
\end{align}
\end{subequations}
where $g^{\theta_i}_s \ge 0$ is the number of molecules of type $\theta_i$ allocated to Tx-s and $\Lambda$ is the budget of all types of molecules allocated to each transmitter node. Furthermore, $\textbf{G}=[g^{\theta_i}_s]_{1\leq i\leq r,1\leq s\leq n}$. As stated in \cite{cao2019diffusive}, the optimization problem~(\ref{eq:STDN}) is convex. Note that we assume the number of molecules as real variable in (\ref{eq:STDN}). After optimizing the number of molecules, we quantize them to the nearest integer value. However, the difference between considering $\textbf{G}$ as real and integer variables is discussed in the next section. To solve the optimization problem~\ref{eq:STDN}, we utilize publicly available software CVX~\cite{grant2008cvx}.

\section{Numerical Result}\label{sec:numerical_result}
In this section, we provide the numerical analysis of the proposed MCvD system. We consider four types of amino-acids as $\theta_1=\text{Glysine}$ with $D_{\theta_1} = \text{10.40}\times\text{10}^{\text{-10}}$, $\theta_2=\text{L-Alanine}$ with $D_{\theta_2} = \text{9.04}\times\text{10}^{\text{-10}}$, $\theta_3=\beta\text{-Alanine}$ with $D_{\theta_3} = \text{9.36}\times\text{10}^{\text{-10}}$, and $\theta_4=\text{L-Serine}$ with $D_{\theta_4} = \text{9.16}\times\text{10}^{\text{-10}}$ as the messenger molecules in the Aqueous diffusive environment~\cite{ma2005studies}. We utilize four transmitters and four receivers, i,e., $n=\text{4}$ and $r = \text{4}$. The transmitters are placed in a column with $\text{2}\mu\text{m}$ gap between them and the receivers are distanced $\text{25}\mu\text{m}$ from transmitters. The radius of the receivers are considered as $\text{7}\mu\text{m}$. In addition, we consider 10 previous time slots to calculate ISI, due to the fact that the ISI length is limited~\cite{tavakkoli2017performance}.

The BER of the MCvD system for SISO, SIMO, MISO, MIMO-STSR, and MIMO-MTMR are illustrated in Fig.~\ref{Fig:BER}. We utilize the Practical Zero Forcing~(PZF) method in detection block for MIMO-STSR~\cite{koo2015detection}, and use molecules of type~$\theta_1$ to send information. We allocate equal number of molecules to each transmitter as 1000 for each SISO, SIMO, MISO, and MIMO frameworks to compare them fairly. Fig.~\ref{Fig:BER} shows that the considered MISO performance is better than that of SISO, SIMO, MIMO-STSR, and MIMO-MTMR, due to the fact that the transmitters release one type of molecules to the medium. In spite of that, the bit rate~($r/t$) of MISO is equal to SISO and lower than that of SIMO and MIMO. By considering the bit rate and BER jointly, MIMO-MTMR has the best performance among others, because its BER is better than SIMO. It is worth noting that MIMO-MTMR outperforms MIMO-STSR, because MIMO-STSR use just one type of molecules to transmit information. By taking all the explanations into account, if the bit rate is not important in the considered application, we suggest the proposed MISO as the best one. But, if the considered application suggests both the BER and bit rate, MIMO-MTMR is better than the others in the proposed scheme for $t > \text{2}\text{s}$. However, for $t=\text{1}\text{s}$, Fig.~\ref{Fig:BER} shows that MIMO-STSR is better than the others. But, for larger values of time slots, MIMO-MTMR is better than MIMO-STSR.
 The BER of MIMO-MTMR reaches to $\text{3.6}\times\text{10}^{\text{-2}}$ for~$t = \text{10s}$ while its bit rate is 0.4, but, in spite of that, the BER of MISO reaches to $\text{2.2}\times\text{10}^{\text{-2}}$ for $t = \text{10s}$ and its bit rate is 0.1 where is lower than that of MIMO-MTMR.

In Fig.~\ref{Fig:Opt_BER}, we illustrate the minimized BER as a function of time slot by optimizing the number of molecules allocated to each transmitter in case of utilizing MIMO-MTMR MCvD system. It shows that by increasing the budget of molecules for each transmitter, the BER gets better. By setting $t = \text{10s}$, the BER is minimized to $\text{1.4}\times\text{10}^{\text{-2}}$ and $\text{3.7}\times\text{10}^{\text{-3}}$ for the budget of molecules considered as 1000 and 10000, respectively. Fig.~\ref{Fig:Opt_BER} also shows that considering the number of molecules' variable as real or integer has very little difference in the objective function. The BER difference between considering number of molecules as real and integer for $\Lambda=\text{50}$ is $\text{4.9}\times\text{10}^{\text{-3}}$. However, for $\Lambda=\text{10000}$, the BER difference is $\text{3}\times\text{10}^{\text{-5}}$. Thus, by increasing the number of molecules allocated to each transmitter, the BER difference between considering the number of molecules' variable as real and integer is decreased considerably. Therefore, we can state that considering $\textbf{G}$ as a real variable in the optimization problem (\ref{eq:STDN}) does not change the solution. It is worth noting that by quantizing the number of molecules variable to the nearest integer value, the budget of molecules might exceeds a few number of molecules. However, this problem does not affect the performance of the system.

\section{Conclusion and Future Works}\label{sec:conclusion}
In this paper, we proposed a new framework for MIMO-based MCvD systems utilizing different types of molecules to convey information as MTMR. The proposed MIMO-MTMR MCvD can ba applied in DDSs to deliver different drugs to different target cells. We also investigated the performance of the system for SISO, SIMO, MISO, and MIMO-STSR cases. An optimization problem is also formulated to optimize the number of molecules, i.e., the drug dosage in DDS, to minimize the error probability and design the drug releasing mechanism. The numerical results showed that by considering the bit rate and BER of the proposed MCvD system, MIMO-MTMR is better than other cases~(SISO, SIMO, MISO, and MIMO-STSR). 

As future works, we aim to investigate the proposed MCvD system for designing the drug release mechanism in mobile MC to deliver drugs toward the cancer cells.

\bibliographystyle{ieeetr}
\bibliography{Hamid_Khoshfekr}

\end{document}